\documentclass[twocolumn,english,nofootinbib]{revtex4-2}
\usepackage[T1]{fontenc}
\usepackage[latin9]{inputenc}
\usepackage[a4paper]{geometry}
\usepackage{amsmath}
\usepackage{amssymb}
\usepackage{cancel}
\usepackage{stmaryrd}
\usepackage[dvipsnames]{xcolor}
\usepackage[normalem]{ulem}
\usepackage{tikz}

\makeatletter
\usepackage{hyperref}
\usepackage{slashed}
\usepackage{shuffle}

\makeatother

\usepackage{babel}
\begin{document}
\title{Four-dimensional spectrum generating algebra for the superstring}
\author{Renann Lipinski Jusinskas$^{a}$}
\author{Aravindhan Srinivasan$^{b,c}$}

\affiliation{$^{a}$ Institute of Physics of the Czech Academy of Sciences \& CEICO
~\\
Na Slovance 2, 18221 Prague, Czech Republic}
\affiliation{$^{b}$Institute of Mathematics, Czech Academy of Sciences,~\\
\v Zitn\' a 25, 115 67 Prague 1, Czech Republic,}
\affiliation{$^{c}$Institute of Theoretical Physics, Faculty of Mathematics and
Physics, ~\\
Charles University, V Hole\v{s}ovi\v{c}k\'{a}ch 2, 180 00 Prague 8, Czech Republic.}

\date{}
\allowdisplaybreaks

\begin{abstract}
We present the spectrum generating algebra for the hybrid string
with manifest $\mathcal{N}=1$ super Poincar\'e symmetry in $\mathbb{R}^{3,1}\times\mathbb{R}^{6}$.
Our DDF operators establish a one-to-one correspondence with the conventional
superstring spectrum, while making its four-dimensional structure
manifest. We also discuss part of the antifield spectrum, and introduce
a simple realization of the supersymmetry charges involving the $\mathbb{R}^{6}$
directions. As an application, we algebraically compute the helicity
partition function. These results provide new tools for analyzing
four-dimensional on-shell superspaces and may extend naturally to
phenomenologically relevant compactifications.
\end{abstract}
\maketitle

\section{Introduction}

String theory stands out as a self-consistent framework incorporating gravity at the quantum level. Despite this achievement, a string vacuum that precisely
matches the matter content and parameters of our universe has not
yet been identified. Among other notable features, string theory requires
additional spacetime dimensions; these are typically argued to be
Planck-scale small or part of large extra dimension scenarios (see
\cite{Arkani-Hamed:1998jmv} and \cite{Randall:1999ee}). Their precise
geometry and size determine the four-dimensional physics, contributing
to the vast string landscape of possible vacua (see \cite{Agmon:2022thq}
for a recent review). Experimental constraints are stringent but have
not yet ruled out their existence, in particular because their nature
is highly model-dependent \cite{ParticleDataGroup:2022pth}. From
a phenomenological perspective, it is then natural to try to formulate
string theory in a way that makes four-dimensional physics manifest.
This can be achieved to different degrees of convenience through compactification
in the more traditional formalisms (e.g. \cite{Feng:2012bb}), though
not without technical challenges.

In \cite{Berkovits:1994wr,Berkovits:1996bf}, the critical spinning
string was recast in terms of a model in which super Poincar\'e symmetry
is manifest in four spacetime dimensions, with minimal supersymmetry.
This  hybrid string has the desirable features of the Ramond-Neveu-Schwarz
and Green-Schwarz formalisms, while sidestepping certain complicating
ingredients. It has a unique phenomenological appeal, enabling covariant
quantization in $d=4$ without the need for spectrum projections and
sum over spin structures. The hybrid formalism has been studied and
applied in different contexts, in both its $d=4$ and $d=6$ \cite{Berkovits:1994vy}
formulations. Earlier results showed that it is well-suited to describe
string theory in Ramond-Ramond backgrounds \cite{Berkovits:1999im,Berkovits:1999zq,Berkovits:1999xv}.
More recently, it has been used to study the $\text{AdS}_{3}/\text{CFT}_{2}$
correspondence \cite{Eberhardt:2018ouy,Eberhardt:2019ywk}, and second-quantized
heterotic strings \cite{Portugal:2024mvp,Berkovits:2024fpk}, showcasing
its versatility.

Regarding physical states, the covariant construction of vertex operators
in the hybrid string becomes impractical at higher mass levels. In
$d=4$, beyond the massless spectrum only the first massive level
of the open string has been explored -- either in flat space or in
a constant electromagnetic background \cite{Berkovits:1997zd,Benakli:2021jxs}.
The $d=6$ formulation is in this sense more developed (e.g. \cite{Dolan:1999dc,Gaberdiel:2021njm,Daniel:2024ymb}).
Spectrum generating algebras have been successfully obtained in $AdS_{3}$
backgrounds \cite{Giveon:1998ns,Eberhardt:2019qcl,Naderi:2022bus},
extending the Del Giudice--Di Vecchia--Fubini (DDF) construction
\cite{DelGiudice:1971yjh}. Analogous progress in the $d=4$ version
has not yet been reported.

Such a gap becomes particularly substantial given the recent applications
found for DDF operators. They have been used, for example, in the
analysis of the decay and scattering of highly excited strings, and
their chaotic behavior \cite{Gross:2021gsj,Rosenhaus:2021xhm} (see also \cite{Biswas:2024mdu}). The
latter motivated the introduction of a new measure for chaos in scattering
amplitudes \cite{Bianchi:2022mhs,Bianchi:2023uby}. In these results,
the focus was on DDF operators of the bosonic string, with the tachyon
as ground state.
In the spinning string \cite{Green:2012oqa}, the
creation-annihilation algebra of Ramond vertices leads to operators
of arbitrarily high picture number (see e.g. \cite{Aldi:2019osr}).
This hinders a more systematic study of scattering amplitudes. In
the pure spinor superstring this obstacle is naturally absent \cite{Mukhopadhyay:2005zu,Jusinskas:2014vqa,Berkovits:2014bra}.
However, amplitude computations are more convoluted because of the
pure spinor prescription, in particular when brought down to $d=4$.

In this work, we present a spectrum generating algebra for the hybrid
string in $\mathbb{R}^{3,1}\times\mathbb{R}^{6}$, introducing a direct
vertex operator map to more traditional superstring formalisms. In
addition to making manifest the $d=4$ structure of the vertex operators,
our construction bypasses current limitations of DDF operators in
superstring theories. We also derive some simple partition functions
of the spectrum, including the distribution of helicities through
different mass levels. This data is helpful in understanding interactions
and selection rules sensitive to helicity, and the counting of physical
states relevant for helicity amplitudes. More generally, it helps
to expose the structure of four-dimensional on-shell superspaces and
spinor-helicity descriptions in string theory. Furthermore, our results
establish a framework for the algebraic description of the physical
spectrum and the moduli space of superstrings on backgrounds of phenomenological
interest, such as the $T^{6}/Z_{3}$ heterotic orbifold \cite{Dixon:1985jw,Dixon:1986jc}.

\section{The hybrid formalism}

We will split the target space coordinates of the string, $X^{\mu}$,
with $\mu=0,\ldots,9$, into $x^{m}$, with $m=0,\ldots,3$, and $\{\bar{y}^{I},y_{I}\}$,
with $I=1,2,3$. Their world-sheet action is given by
\begin{equation}
S_{X}=\frac{1}{2\pi\alpha'}\int d^{2}z\bigg\{\partial x^{m}\bar{\partial}x^{n}\eta_{mn}+\partial\bar{y}^{I}\bar{\partial}y_{I}\bigg\},
\end{equation}
where $\eta_{mn}$ is the flat metric. The remaining world-sheet fields
are described by the action
\begin{multline}
S_{f}=\frac{1}{2\pi}\int d^{2}z\bigg\{ p_{\alpha}\bar{\partial}\theta^{\alpha}+\bar{p}_{\dot{\alpha}}\bar{\partial}\bar{\theta}^{\dot{\alpha}} \\
+\bar{\psi}^{I}\bar{\partial}\psi_{I}+\frac{1}{2}\partial\rho\bar{\partial}\rho+\ldots\bigg\},
\end{multline}
where $\alpha,\dot{\alpha}=\pm$, with $\theta^{\alpha}$ and $\bar{\theta}^{\dot{\alpha}}$
denoting the spacetime superpartners of $x^{m}$, and $\bar{\psi}^{I}$
and $\psi_{I}$ the twisted world-sheet superpartners of $\bar{y}^{I}$
and $y_{I}$. In addition, we have a chiral scalar $\rho$, satisfying
$\bar{\partial}\rho=0$, and the ellipsis denotes an antiholomorphic
sector. The latter could be the usual mirrored fields of the open
string or heterotic/type II completions of the closed string. For
simplicity, we will focus on the holomorphic sector, and specific
results for open or closed strings follow the standard steps. 

The four-dimensional spacetime supersymmetry charges are realized
through\begin{subequations}\label{eq:susycharges}
\begin{align}
q_{\alpha} & =\oint(p_{\alpha}-\frac{1}{\alpha'}\sigma_{\alpha\dot{\alpha}}^{m}\bar{\theta}^{\dot{\alpha}}\partial x_{m}-\frac{1}{4\alpha'}\partial\theta_{\alpha}\bar{\theta}^{\dot{\alpha}}\bar{\theta}_{\dot{\alpha}}),\\
\bar{q}_{\dot{\alpha}} & =\oint(\bar{p}_{\dot{\alpha}}-\frac{1}{\alpha'}\sigma_{\alpha\dot{\alpha}}^{m}\theta^{\alpha}\partial x_{m}-\frac{1}{4\alpha'}\partial\bar{\theta}_{\dot{\alpha}}\theta^{\alpha}\theta_{\alpha}),
\end{align}
\end{subequations}which satisfy $\{q_{\alpha},q_{\beta}\}=\{\bar{q}_{\dot{\alpha}},\bar{q}_{\dot{\beta}}\}=0$
and
\begin{equation}
\{q_{\alpha},\bar{q}_{\dot{\alpha}}\}=-\frac{2}{\alpha'}\sigma_{\alpha\dot{\alpha}}^{m}\oint\partial x_{m}.
\end{equation}
The Pauli matrices are denoted by $\sigma_{\alpha\dot{\alpha}}^{m}$,
and spinor indices are raised (lowered) with the antisymmetric tensor
$\epsilon^{\alpha\beta}$ ($\epsilon_{\alpha\beta})$, e.g. $\theta_{\alpha}=\epsilon_{\alpha\beta}\theta^{\beta}$,
with $\epsilon^{\alpha\gamma}\epsilon_{\gamma\beta}=\delta_{\beta}^{\alpha}$.
There is an analogous construction for the dotted indices. Unlike
in the spinning string, the $d=4$ supersymmetry algebra in the hybrid formalism
closes without picture changing. 

In order to have manifest supersymmetry, the usual invariants are
introduced,
\begin{align}
d_{\alpha} & =p_{\alpha}+\frac{1}{2\alpha'}\bar{\theta}^{\dot{\alpha}}(2\sigma_{\alpha\dot{\alpha}}^{m}\partial x_{m}+\bar{\theta}_{\dot{\alpha}}\partial\theta_{\alpha}-\partial\bar{\theta}_{\dot{\alpha}}\theta_{\alpha}),\\
\bar{d}_{\dot{\alpha}} & =\bar{p}_{\dot{\alpha}}+\frac{1}{2\alpha'}\theta^{\alpha}(2\sigma_{\alpha\dot{\alpha}}^{m}\partial x_{m}+\theta_{\alpha}\partial\bar{\theta}_{\dot{\alpha}}-\partial\theta_{\alpha}\bar{\theta}_{\dot{\alpha}}),\\
\Pi^{m} & =\partial x^{m}+\frac{1}{2}\sigma_{\alpha\dot{\alpha}}^{m}(\partial\theta^{\alpha}\bar{\theta}^{\dot{\alpha}}+\partial\bar{\theta}^{\dot{\alpha}}\theta^{\alpha}).
\end{align}
The hybrid string BRST current is given by
\begin{equation}
J_{\textrm{BRST}}=e^{\rho}\Delta+\frac{4}{\alpha'}\partial\bar{y}^{I}\psi_{I},
\end{equation}
while the $b$ ghost, due to various field redefinitions, is recast
as
\begin{equation}
b=-\frac{\alpha'}{16}e^{-\rho}\bar{\Delta}-\frac{1}{4}\partial y_{I}\bar{\psi}^{I},
\end{equation}
where
\begin{align}
\Delta & = \epsilon^{\alpha\beta}(d_{\alpha},d_{\beta}),\\
\bar{\Delta} & = \epsilon^{\dot{\alpha}\dot{\beta}}(\bar{d}_{\dot{\alpha}},\bar{d}_{\dot{\beta}}).
\end{align}
The ordering operation is defined by
\begin{equation}
(A,B)(z)=\oint\frac{dw}{(w-z)}A(w)B(z).
\end{equation}

The holomorphic component of the total energy-momentum tensor,  given by
\begin{multline}
T=-\frac{1}{\alpha'}\partial x^{m}\partial x_{m}-p_{\alpha}\partial\theta^{\alpha}-\bar{p}_{\dot{\alpha}}\partial\bar{\theta}^{\dot{\alpha}}\\
-\frac{1}{\alpha'}\partial\bar{y}^{I}\partial y_{I}-\bar{\psi}^{I}\partial\psi_{I}-\frac{1}{2}\partial\rho\partial\rho-\frac{1}{2}\partial^{2}\rho,\label{eq:EMtensor}
\end{multline}
is BRST-exact. This follows from the OPE
\[
J_{\textrm{BRST}}(z)\,b(y)\sim\frac{2}{(z-y)^{3}}+\frac{J}{(z-y)^{2}}+\frac{T}{(z-y)}.
\]
Here, $J$ is a $U(1)$ current that is identified as the ghost number,
and expressed as
\begin{equation}
J=\psi_{I}\bar{\psi}^{I}-\partial\rho.
\end{equation}

Lastly, we introduce
\begin{equation}
\eta=-2\alpha'e^{-2\rho}\Psi\bar{\Delta}+4\epsilon^{IJK}e^{-\rho}\partial y_{I}\psi_{J}\psi_{K},
\end{equation}
with $\epsilon^{IJK}$ denoting the totally antisymmetric tensor with
$\epsilon^{123}=+1$, and $\Psi=(\epsilon^{IJK}\psi_{I}\psi_{J}\psi_{K})/6$.
The operator $\eta$ comes from the fermionization of the superconformal
ghosts of the spinning string, again after field redefinitions and
similarity transformations.

\section{The massless spectrum}

Physical states $U$ in the hybrid string belong to the ghost number
one cohomology of the BRST charge
\begin{equation}
Q=\oint J_{\textrm{BRST}}.
\end{equation}
Furthermore, the physical spectrum should be in the so-called small
Hilbert space. This condition is inherited from the spinning string,
and can be expressed as $\eta_{0}\cdot U=0$, i.e., physical states
are also annihilated by the zero mode of $\eta$.

Since our goal is to have manifest Lorentz symmetry in four dimensions,
we will boost all momentum eigenstates to the $x^{m}$ directions.
In this case, the massless spectrum can be divided into two sets.
The first corresponds to states that are singlets with respect to
boosts in the $\{\bar{y}^{I},y_{I}\}$ directions:
\begin{multline}
U_{\textrm{4d}} = \eta_{0}\cdot\mathcal{V}(x^{m},\theta^{\alpha},\bar{\theta}^{\dot{\alpha}}), \\
=-2\alpha'[\partial(\Psi e^{-2\rho})\bar{D}^{2}\mathcal{V}+2\Psi e^{-2\rho}(\bar{d}_{\dot{\alpha}},\bar{D}^{\dot{\alpha}}\mathcal{V})].\label{eq:4dun}
\end{multline}
Here $\mathcal{V}$ is a real superfield and $\bar{D}_{\dot{\alpha}}$
($D_{\alpha}$) denotes the superderivative, with $\bar{D}^{2}=\bar{D}_{\dot{\alpha}}\bar{D}^{\dot{\alpha}}$.
BRST-closedness requires that
\begin{equation}
D_{\alpha}\bar{D}^{2}D^{\alpha}\mathcal{V}=\bar{D}_{\dot{\alpha}}D^{2}\bar{D}^{\dot{\alpha}}\mathcal{V}=0,
\end{equation}
which is the equation of motion for Maxwell's superfield. The gauge
invariance $\delta\mathcal{V}=\Lambda+\bar{\Lambda},$ where $\Lambda$
($\bar{\Lambda}$) is a chiral (antichiral) superfield, follows directly
from $\eta_{0}$-exact (BRST-exact) states. On the mass shell, $\mathcal{V}$
encodes two bosonic and two fermionic degrees of freedom, respectively
the two circular polarizations of the photon and superpartners. In
the case of closed strings, they would describe the chiral half of
the massless multiplet.

The remaining physical states at the massless level can be mapped
to the vertex operator
\begin{equation}
U_{\textrm{6d}}=\psi_{I}\bar{\Phi}^{I}+e^{-\rho}\epsilon^{IJK}\psi_{J}\psi_{K}\Phi_{I},\label{eq:6dun}
\end{equation}
satisfying the massless superfield equations
\begin{equation}
D^{2}\Phi_{I}=\bar{D}^{2}\bar{\Phi}^{I}=0,
\end{equation}
together with the chirality conditions $D_{\alpha}\bar{\Phi}^{I}=\bar{D}_{\dot{\alpha}}\Phi_{I}=0$.
The massless equation of motion for $\Phi_{I}$ follows from BRST-closedness,
while annihilation by $\eta_{0}$ imply that $\bar{D}_{\dot{\alpha}}\Phi_{I}=0$.
For $\bar{\Phi}_{I}$, it is the other way around, with $Q$ and $\eta_{0}$
playing a dual role. In fact, $\bar{\Phi}^{I}$ and $\Phi_{I}$ are
conjugate to each other in the hybrid formalism\footnote{Hermiticity properties have a non-trivial realization in the hybrid
description \cite{Berkovits:1996cc}.}. All together, they describe six bosonic and six fermionic degrees
of freedom. In \cite{Berkovits:1994wr} there is a detailed analysis
of the massless cohomology in the heterotic case in a Calabi-Yau background.

It is interesting to point out that the hybrid string also inherits
the picture degeneracy of the spinning string. Therefore, a given
physical state admits multiple (equivalent) descriptions with vertex
operators at different pictures. For example, we can define $\tilde{U}_{\textrm{6d}}=\tfrac{1}{2}\eta_{0}\cdot(e^{-\rho}\psi_{I}\Omega^{I})$,
such that
\begin{multline}
\tilde{U}_{\textrm{6d}}=4e^{-2\rho}\Psi\partial y_{I}\Omega^{I}-\alpha'e^{-3\rho}\Psi\partial\psi_{I}(\bar{\Delta},\Omega^{I})\\
+\alpha'e^{-3\rho}(\partial^{2}\Psi-4\partial\Psi\partial\rho)\psi_{I}(\bar{d}_{\dot{\alpha}},\bar{D}^{\dot{\alpha}}\Omega^{I})\\
+\alpha'[\partial\Psi\partial^{2}\rho-2\partial\Psi(\partial\rho)^{2}]e^{-3\rho}\psi_{I}\bar{D}^{2}\Omega^{I}\\
+\alpha'(\partial^{2}\Psi\partial\rho-\tfrac{1}{3!}\partial^{3}\Psi)e^{-3\rho}\psi_{I}\bar{D}^{2}\Omega^{I}.
\end{multline}
It is then straightforward to evaluate the action of the BRST charge,
\begin{equation}
Q\cdot\tilde{U}_{\textrm{6d}}=-\alpha'e^{-2\rho}\Psi\partial\psi_{I}\bar{D}^{2}D^{2}\Omega^{I}.
\end{equation}
In this case, $D^{2}\Omega^{I}$ defines the anti-chiral superfield
$\bar{\Phi}^{I}$, and BRST closedness leads to $\bar{D}^{2}\bar{\Phi}^{I}=0$.

The vertices (\ref{eq:4dun}) and (\ref{eq:6dun}) also have the usual
integrated version. The respective integrands are
\begin{multline}
V_{\textrm{4d}}=\mathrm{i}\partial\theta^{\alpha}D_{\alpha}\mathcal{V}-\mathrm{i}\partial\bar{\theta}^{\dot{\alpha}}\bar{D}_{\dot{\alpha}}\mathcal{V}-\frac{\mathrm{i}\alpha'}{4}(d_{\alpha},\bar{D}^{2}D^{\alpha}\mathcal{V})\\
+\frac{\mathrm{i}\alpha'}{4}(\bar{d}_{\dot{\alpha}},D^{2}\bar{D}^{\dot{\alpha}}\mathcal{V})+\frac{\mathrm{i}}{2}\sigma_{\alpha\dot{\alpha}}^{m}(\Pi_{m},[\bar{D}^{\dot{\alpha}},D^{\alpha}]\mathcal{V}),\label{eq:4din}
\end{multline}
in which $\delta\mathcal{V}=\Lambda+\bar{\Lambda}$ leads to $\delta V_{\textrm{4d}}=\mathrm{i}\partial(\Lambda-\bar{\Lambda})$,
and
\begin{multline}
V_{\textrm{6d}}=\partial\bar{y}^{I}\Phi_{I}+\frac{\alpha'}{2}e^{\rho}\bar{\psi}^{I}(d_{\alpha},D^{\alpha}\Phi_{I})\\
+\partial y_{I}\bar{\Phi}^{I}+\frac{\alpha'}{2}e^{-\rho}\psi_{I}(\bar{d}_{\dot{\alpha}},\bar{D}^{\dot{\alpha}}\bar{\Phi}^{I}).\label{eq:6din}
\end{multline}
The latter should be equivalent to the large radius limit of marginal
deformations of the Calabi-Yau manifold, which were described in \cite{Berkovits:1994wr}.
Both $V_{\textrm{4d}}$ and $V_{\textrm{6d}}$ are annihilated by
the BRST charge and by $\eta_{0}$ up to total derivatives. For example,
we can show that
\begin{eqnarray}
Q\cdot V_{\textrm{4d}} & = & \partial[\partial e^{\rho}D^{2}\mathcal{V}+2e^{\rho}(d_{\alpha},D^{\alpha}\mathcal{V})],\\
\eta_{0}\cdot V_{\textrm{6d}} & = & -4\alpha'\partial[e^{-\rho}\epsilon^{IJK}\psi_{J}\psi_{K}\Phi_{I}],
\end{eqnarray}
using the relevant superfield equations of motion.

Note that from (\ref{eq:6din}), we obtain the missing supersymmetry
charges of the model. They are given by\begin{subequations}\label{eq:susycharges-6d}
\begin{eqnarray}
q_{\alpha}^{I} & = & \oint(e^{\rho}\bar{\psi}^{I}d_{\alpha}-\frac{2}{\alpha'}\partial\bar{y}^{I}\theta_{\alpha}),\\
\bar{q}_{I\dot{\alpha}} & = & \oint(e^{-\rho}\psi_{I}\bar{d}_{\dot{\alpha}}-\frac{2}{\alpha'}\partial y_{I}\bar{\theta}_{\dot{\alpha}}),
\end{eqnarray}
\end{subequations}and satisfy
\begin{equation}
\{q_{\alpha}^{I},\bar{q}_{J\dot{\alpha}}\}=-\frac{2}{\alpha'}\delta_{J}^{I}\sigma_{\alpha\dot{\alpha}}^{m}\oint\partial x_{m}.
\end{equation}
The remaining brackets, including with $q_{\alpha}$ and $\bar{q}_{\dot{\alpha}}$,
are either BRST-exact or $\eta_{0}$-exact. This is compatible with
the fact that the $\mathcal{N}=1$ ten-dimensional supersymmetry algebra
only closes on-shell. It is straightforward to show that (\ref{eq:4dun})
and (\ref{eq:6dun}) are mapped to each other using (\ref{eq:susycharges-6d}). 

As for the massless antifield spectrum, i.e., the ghost number two
cohomology, it is easy to find the analogue of (\ref{eq:4dun}),
for instance. A generic $\eta_{0}$-exact vertex operator can be cast
as
\begin{equation}
U_{4d}^{*}=e^{-\rho}\Psi\Phi^{*}(x,\theta,\bar{\theta}),
\end{equation}
with $\Phi^{*}\equiv\bar{D}_{\dot{\alpha}}\bar{A}^{\dot{\alpha}}$
for a given superfield $\bar{A}^{\dot{\alpha}}$. BRST-closedness
implies the equation of motion $D^{2}\Phi^{*}=0$. BRST-exact states
translate to the gauge invariance $\delta\bar{A}^{\dot{\alpha}}=\bar{\Lambda}^{\dot{\alpha}}$,
with $\bar{\Lambda}^{\dot{\alpha}}$ denoting an antichiral superfield
parameter. In order to manifest the field content of antifield multiplet,
we can fix some algebraic redundancy, such that
\begin{multline}
\Phi^{*}=\theta^{\alpha}\chi_{\alpha}^{*}+\theta^{\alpha}\theta_{\alpha}\bar{\theta}^{\dot{\alpha}}\xi_{\dot{\alpha}}^{*}+\theta^{\alpha}\theta_{\alpha}F^{*}\\
+\theta^{\alpha}\bar{\theta}^{\dot{\alpha}}\sigma_{\alpha\dot{\alpha}}^{m}a_{m}^{*}-\frac{1}{4}\theta^{\alpha}\theta_{\alpha}\bar{\theta}^{\dot{\alpha}}\bar{\theta}_{\dot{\alpha}}\partial^{m}a_{m}^{*}.
\end{multline}
$D^{2}\Phi^{*}=0$ then leads to\begin{subequations}
\begin{eqnarray}
F^{*} & = & 0,\\
\partial^{m}a_{m}^{*} & = & 0,\\
\xi_{\dot{\alpha}}^{*} & = & \frac{1}{4}\epsilon^{\alpha\beta}\sigma_{\alpha\dot{\alpha}}^{m}\partial_{m}\chi_{\beta}^{*},
\end{eqnarray}
\end{subequations}matching the expected antifield dynamics of the
massless vector multiplet.

\section{Spectrum generating algebra}

While the analysis of the massless spectrum is relatively simple,
the computation of massive string vertices quickly becomes impractical.
Results for the first massive level can be found in \cite{Berkovits:1997zd,Benakli:2021jxs}.
For higher mass levels, there is a systematic procedure to build vertex
operators of arbitrary mass level using only the massless ones, the
well-known DDF operators \cite{DelGiudice:1971yjh}. Simply put, they
correspond to massless vertices  rewritten in a specific
light-cone frame.

We will denote the four-dimensional directions $x^{m}$ by $x^{\pm}=(x^{0}\pm x^{3})/\sqrt{2}$,
$x=(x^{1}+\mathrm{i}x^{2})/\sqrt{2}$, and $\bar{x}=(x^{1}-\mathrm{i}x^{2})/\sqrt{2}$,
such that the non-zero components of the flat-space metric are $\eta^{+-}=\eta^{-+}=-1$
and $\eta^{x\bar{x}}=\eta^{\bar{x}x}=1$. In this case, the Pauli
matrices have a simple form, with $\sigma_{\alpha\dot{\alpha}}^{\pm}=\sqrt{2}\delta_{\alpha}^{\pm}\delta_{\dot{\alpha}}^{\pm}$,
$\sigma_{\alpha\dot{\alpha}}^{x}=\sqrt{2}\delta_{\alpha}^{-}\delta_{\dot{\alpha}}^{+}$,
and $\sigma_{\alpha\dot{\alpha}}^{\bar{x}}=\sqrt{2}\delta_{\alpha}^{+}\delta_{\dot{\alpha}}^{-}$.

Our first step is to introduce the ground states of the DDF representation.
They will be defined through the state-operator map of the gauge-fixed
massless unintegrated vertex operators of (\ref{eq:4dun}) and (\ref{eq:6dun})
with momentum $\sqrt{2}k$ in the light-cone direction $x^{-}$. Up
to normalizations, the bosonic states are identified by\begin{subequations}\label{eq:groundstates}
\begin{eqnarray}
\left|a,k\right\rangle  & \cong & \Psi e^{-2\rho}\bar{d}_{+}\theta^{+}e^{\mathrm{i}k(\sqrt{2}X^{-}-\theta^{-}\bar{\theta}^{-})},\\
\left|\bar{a},k\right\rangle  & \cong & \Psi e^{-2\rho}\bar{d}_{-}\theta^{-}e^{\mathrm{i}k\sqrt{2}X^{-}},\\
\left|b,I,k\right\rangle  & \cong & e^{-\rho}\epsilon^{IJK}\psi_{J}\psi_{K}e^{\mathrm{i}k(\sqrt{2}X^{-}-\theta^{-}\bar{\theta}^{-})},\\
\left|\bar{b},I,k\right\rangle  & \cong & \psi_{I}e^{\mathrm{i}k(\sqrt{2}X^{-}+\theta^{-}\bar{\theta}^{-})},
\end{eqnarray}
and the fermionic states by
\begin{eqnarray}
\left|\chi,k\right\rangle  & \cong & -2\mathrm{i}\Psi e^{-2\rho}\bar{d}_{+}\theta^{+}\theta^{-}e^{\mathrm{i}k\sqrt{2}X^{-}},\\
\left|\bar{\chi},k\right\rangle  & \cong & \frac{1}{k}\Psi e^{-2\rho}\bar{d}_{-}e^{\mathrm{i}k(\sqrt{2}X^{-}+\theta^{-}\bar{\theta}^{-})},\\
\left|\xi,I,k\right\rangle  & \cong & e^{-\rho}\epsilon^{IJK}\psi_{J}\psi_{K}\theta^{-}e^{\mathrm{i}k\sqrt{2}X^{-}},\\
\left|\bar{\xi},I,k\right\rangle  & \cong & \psi_{I}\bar{\theta}^{-}e^{\mathrm{i}k\sqrt{2}X^{-}}.
\end{eqnarray}
\end{subequations}For convenience, these states will be collectively
denoted by $\left|0,k\right\rangle $, indicating they are massless.
It is straightforward to compute their supersymmetry transformations
using (\ref{eq:susycharges}), e.g. $q_{-}\left|\bar{a},k\right\rangle =k\left|\bar{\chi},k\right\rangle $
and $\bar{q}_{-}\left|\bar{\xi},I\right\rangle =-\left|\bar{b},I\right\rangle $.
In particular, it is possible to show that both $q_{+}\left|a,k\right\rangle $
and $q_{+}\left|\chi,k\right\rangle $ are mapped to pure gauge operators.

For the DDF algebra, we will work with integrated vertex operators
derived from (\ref{eq:4din}) and (\ref{eq:6din}), with momentum
$\sqrt{2}k$ in the light-cone direction $x^{+}$. They are given
by\begin{subequations}\label{eq:DDF-operators}
\begin{align}
A(k) & =\oint[\sqrt{2}\Pi^{x}\Phi-(\frac{\alpha'}{2}\bar{d}_{-}+\frac{\mathrm{i}}{k}\partial\theta^{-})D_{+}\Phi],\\
\bar{A}(k) & =\oint[\sqrt{2}\Pi^{\bar{x}}\bar{\Phi}-(\frac{\alpha'}{2}d_{-}+\frac{\mathrm{i}}{k}\partial\bar{\theta}^{-})\bar{D}_{+}\bar{\Phi}],\\
B(k) & =\oint[\partial\bar{y}^{I}\Phi_{I}+\frac{\alpha'}{2}e^{\rho}\bar{\psi}^{I}(d_{-},D_{+}\Phi_{I})],\\
\bar{B}(k) & =\oint[\partial y_{I}\bar{\Phi}^{I}+\frac{\alpha'}{2}e^{-\rho}\psi_{I}(\bar{d}_{-},\bar{D}_{+}\bar{\Phi}^{I})],
\end{align}
\end{subequations}with\begin{subequations}
\begin{align}
\Phi & =(\phi+\theta^{+}\chi)e^{\mathrm{i}k(\sqrt{2}X^{+}-\theta^{+}\bar{\theta}^{+})},\\
\bar{\Phi} & =(\bar{\phi}+\bar{\theta}^{+}\bar{\chi})e^{\mathrm{i}k(\sqrt{2}X^{+}+\theta^{+}\bar{\theta}^{+})}\\
\Phi_{I} & =(\phi_{I}+\theta^{+}\chi_{I})e^{\mathrm{i}k(\sqrt{2}X^{+}-\theta^{+}\bar{\theta}^{+})},\\
\bar{\Phi}^{I} & =(\bar{\phi}^{I}+\bar{\theta}^{+}\bar{\chi}^{I})e^{\mathrm{i}k(\sqrt{2}X^{+}+\theta^{+}\bar{\theta}^{+})},
\end{align}
\end{subequations}in which $\{\phi,\bar{\phi},\phi_{I},\bar{\phi}^{I}\}$
and $\{\chi,\bar{\chi},\chi_{I},\bar{\chi}^{I}\}$ respectively denote
the set of bosonic and fermionic polarizations. These operators satisfy\begin{subequations}\label{eq:DDF-algebra}
\begin{multline}
[\bar{A}(k),A(\tilde{k})] = \\-\mathrm{i}k\sqrt{2}\alpha'\delta_{k+\tilde{k}}\left(\phi\bar{\phi}+\frac{i}{2k}\chi\bar{\chi}\right)\oint\partial x^{+},
\end{multline}
\begin{multline}
[\bar{B}(k),B(\tilde{k})] = \\-\mathrm{i}k\sqrt{2}\alpha'\delta_{k+\tilde{k}}\left(\phi_{I}\bar{\phi}^{I}+\frac{i}{2k}\chi_{I}\bar{\chi}^{I}\right)\oint\partial x^{+},
\end{multline}
\end{subequations}and the remaining commutators vanish.

Equation (\ref{eq:DDF-algebra}) can be interpreted as a creation-annihilation
algebra for ground states (\ref{eq:groundstates}) with specific momentum.
Schematically, any mass level $N$ state in the physical spectrum
of the hybrid string can be expressed as
\begin{multline}
\left|N,\{p,\bar{p},q,\bar{q}\}\right\rangle =\prod_{n=1}^{L}\left(A(nk)\right)^{p_{n}}\left(\bar{A}(nk)\right)^{\bar{p}_{n}}\\ \times
\left(B(nk)\right)^{q_{n}}\left(\bar{B}(nk)\right)^{\bar{q}_{n}}\left|0,\frac{1}{\alpha'k}\right\rangle ,\label{eq:physicalvertex}
\end{multline}
with mass $m^{2}=4N/\alpha'$. Here, $p_{n}$, $\bar{p}_{n}$, $q_{n}$,
and $\bar{q}_{n}$ are non-negative integers, such that
\begin{equation}
\sum_{n=1}^{L}n(p_{n}+\bar{p}_{n}+q_{n}+\bar{q}_{n})=N.
\end{equation}
Annihilation operators come with negative $n$, e.g.
\begin{equation}
A(-k)\left|0,\frac{1}{\alpha'k}\right\rangle =0.
\end{equation}
Since we have sixteen DDF ground states, c.f. (\ref{eq:groundstates}),
and sixteen creation operators for each $n$, half bosonic and half
fermionic, it is trivial to establish a one-to-one map between the
light-cone states of the hybrid string and other superstring formalisms.

\section{Some partition functions}

From the DDF algebra, we can easily compute some of the partition
functions of the hybrid string. They are defined as the weighted sum
over the states in the physical Hilbert space, $\mathcal{H}_{\textrm{phys}}$, which
is spanned by (\ref{eq:physicalvertex}). While these are connected
to the one-loop vacuum amplitude of the theory (see e.g. \cite{Berkovits:2001nv}),
we are mostly interested in keeping track of some quantum numbers,
in particular because the DDF states have fixed momentum. We will
use the open string as a case study, but the extension to the closed
string is mostly obvious.

The mass distribution of the physical spectrum can be read from
\begin{eqnarray}
Z(q) & = & \textrm{Tr}_{\mathcal{H}_{\textrm{phys}}}\left(q^{L_{0}}\right),\label{eq:trace}\\
 & = & 16\prod_{n=1}^{\infty}\left(\frac{1+q^{n}}{1-q^{n}}\right)^{8},
\end{eqnarray}
which is a well-known result, where $L_{0}$ is the mass-level operator.
If we insert the fermion number operator, $(-1)^{F}$, i.e., the supertrace
in (\ref{eq:trace}), the result vanishes because supersymmetry is
not broken.

Since we are working with the little group $SO(2)\cong U(1)$, the
helicity structure of the DDF states is also very transparent. In
(\ref{eq:DDF-operators}), the polarizations $\{\phi,\bar{\phi},\chi,\bar{\chi}\}$
can be assigned helicity $\{+1,-1,+\tfrac{1}{2},-\tfrac{1}{2}\}$, while the $R$-charged
polarizations $\{\phi_{I},\bar{\phi}^{I},\chi_{I},\bar{\chi}^{I}\}$
have helicity $\{0,0,-\tfrac{1}{2},+\tfrac{1}{2}\}$. The helicity partition function
is defined as
\begin{equation}
H(q,y)=\textrm{Tr}_{\mathcal{H}_{\textrm{phys}}}\left(q^{L_{0}}y^{2h}\right),\label{eq:Zhel}
\end{equation}
where $h$ is the helicity operator. It is straightforward to show
that
\begin{multline}
H(q,y)=\frac{(1+y)^{4}}{y^{2}}\\
\prod_{n=1}^{\infty}\left(\frac{1}{1-q^{n}}\right)^{6}\frac{(1+q^{n}y)^{4}}{1-q^{n}y^{2}}\frac{(1+q^{n}/y)^{4}}{1-q^{n}/y^{2}},
\end{multline}
which explicitly enumerates supermultiplet content in terms of helicities.
The supertrace version of (\ref{eq:Zhel}) is given by $H(q,-y)$.

There are other possible refinements, such as the inclusion of the
R-charge, but we will not discuss them here. They become more interesting
with additional spacetime structures (such as orbifolds) or compactifications
that break supersymmetry. A detailed account of refined partition
functions can be found in \cite{Lust:2012zv}, where the open superstring
is analyzed in even-dimensional compactifications.

\section{Final remarks}

The DDF framework in string theory exposes a clear map between vertex operators and light-cone states. Here we have established this construction for the hybrid string in $\mathbb{R}^{3,1}\times\mathbb{R}^{6}$. The advantage of the hybrid model over other formalisms is the embedded $d=4$ superspace realization of the vertex operators. Our construction is well-placed to extend recent developments on bosonic DDF operators  \cite{Gross:2021gsj,Rosenhaus:2021xhm,Bianchi:2022mhs,Bianchi:2023uby} to the superstring.

More intriguingly, our results may help to bridge four-dimensional methods for field-theory scattering amplitudes with string scattering. The DDF operators defined in \eqref{eq:DDF-operators} admit a semi-covariant formulation using a reference vector, analogous to the spinor-helicity formalism. In particular, massive vertex operators naturally involve two such vectors -- one associated with the ground state and another with the creation operators -- mirroring the structure of massive spinor-helicity \cite{Arkani-Hamed:2017jhn}.

Regarding non-trivial backgrounds, our results extend naturally
to $\mathbb{R}^{3,1}\times T^{6}$, with the exception of compactification modes. Such states have non-zero (discrete) momentum
in the $T^{6}$ directions, as well as winding number. A relevant
future investigation is to determine how the states in the Narain
lattice fit the framework of DDF operators. To our
knowledge, this topic has received only partial treatment,
without a unified spectrum\nobreakdash-generating algebra in most
of the literature. Perhaps more interestingly, we would like to analyze a similar construction
 in the heterotic orbifold $T^{6}/Z_{3}$. In this
background, the string spectrum  resembles the minimal supersymmetric
standard model, with chiral fermions and a realistic gauge group,
while remaining fully accessible through the powerful methods of free
conformal field theory. The explicit construction of the
vertex operators requires the use of twist fields \cite{Dixon:1985jw,Dixon:1986jc},
and the hybrid formalism may offer a natural framework for this, particularly  the massless sector.

Beyond the spectrum generating algebra, it would be interesting to
explore the hybrid formalism in the context of chiral strings \cite{Hohm:2013jaa,Siegel:2015axg}.
Its tensionless limit should be connected to some variant of the ambitwistor
string \cite{Mason:2013sva}, with four-dimensional covariance. Since the Virasoro
constraints are not manifest in the hybrid string, we expect additional ingredients to play a role. A possible connection to twistor
strings \cite{Witten:2003nn,Berkovits:2004hg} might become more evident in this case. Furthermore,
this investigation might help to introduce alternative Cachazo-He-Yuan
formulae \cite{Cachazo:2013hca} for the scattering of gluons and gravitons, now in superspace.
\begin{acknowledgments}
We thank Ulisses Portugal for useful discussions. A.S. is supported by the GA\v{C}R grant 25-15544S from the Czech Science Foundation, the Charles University Research Center Grant No. UNCE24/SCI/016, and the research plan RVO 67985840 of the Institute of Mathematics, Czech Academy of Sciences. RLJ is supported by the GA\v{C}R grant 25-16244S from the Czech Science Foundation.
\end{acknowledgments}

\end{document}